# Radio and Optical Properties of the blazar PKS 1614+051 at z=3.21

Yu. V. Sotnikova,[1,2] A. G. Mikhailov,[1] A. E. Volvach,[3] D. O. Kudryavtsev,[1] T. V. Mufakharov,[1,2] V. V. Vlasyuk,[1] M. L. Khabibullina,[1] A. A. Kudryashova,[1] M. G. Mingaliev[†,1,2] A. K. Erkenov,[1] Yu. A. Kovalev,[4] Y. Y. Kovalev,[5] M. A. Kharinov,[6] T. A. Semenova,[1] R. Yu. Udovitskiy,[1] N. N. Bursov,[1] S. A. Trushkin,[1] O. I. Spiridonova,[1] A. V. Popkov,[7,4] P. G. Tsybulev,[1] L. N. Volvach,[3] N. A. Nizhelsky,[1] G. V. Zhekanis,[1] K. V. Iuzhanina[1,2]

[1]*Special Astrophysical Observatory, Russian Academy of Sciences, Nizhnii Arkhyz, 369167 Russia*
[2]*Kazan (Volga Region) Federal University, Kazan, 420008 Russia*[*]
[3]*Crimean Astrophysical Observatory, Russian Academy of Sciences, Nauchny, 298409 Russia*
[4]*Astro Space Center, Lebedev Physical Institute, Russian Academy of Sciences, Moscow, 117997 Russia*
[5]*Max-Planck-Institut für Radioastronomie, Auf dem Hügel 69, 53121 Bonn, Germany*
[6]*Institute of Applied Astronomy, Russian Academy of Sciences, Kutuzova Embankment 10, St. Petersburg, 191187, Russia*
[7]*Moscow Institute of Physics and Technology, Institutsky per. 9, Dolgoprudny, 141700, Russia*

We present a study of the radio and optical properties of the high-frequency peaker (HFP) blazar PKS 1614+051 at $z = 3.21$ based on the data covering the time period of 1997–2024. The radio data are represented by the almost instantaneous 1–22 GHz measurements from the SAO RAS RATAN-600 radio telescope, the 5 and 8 GHz data from the IAA RAS RT-32 telescopes, and the 37 GHz data from the RT-22 telescope of CrAO RAS. The optical measurements in the $R$ band were collected with the SAO RAS 1-m Zeiss-1000 and 0.5-m AS-500/2 telescopes and the ZTF archive data. We have found low overall variability indices (0.1–0.2) and a spectral peak around 4.6 GHz, which is stable during the long-term period of monitoring. An analysis of the radio light curves reveals significant time delays (0.6 to 6.4 years) between the radio frequencies along with variability timescales ranging from 0.2 to 1.8 years in the source's rest frame, which is similar to the blazars at lower redshifts. Spectral modeling suggests the presence of both synchrotron-self absorption (SSA) and free-free absorption (FFA) processes. Based on the SSA model, we provide estimates of the magnetic field strength, which peaks at $\sim 30$ mG. A spectroscopic study with the BTA SCORPIO-I spectrograph has found signs of the regular motion of a neutral hydrogen envelope around the blazar center, which confirms the presence of enough gaseous matter to form an external FFA screen. The results highlight the importance of multiwavelength and long-term monitoring to understand the physical mechanisms driving the variability in high-redshift blazars.

Ключевые слова: galaxies: active—galaxies: high redshift objects: individual: PKS 1614+051—galaxies: jets—radio continuum: quasars

## 1. INTRODUCTION

Blazars are a subclass of radio-loud quasars with their Doppler-boosted relativistic jets seen at a small angle to the observer's line of sight (Urry and Padovani 1995), and thus they can be visible at high redshifts. High-redshift quasars ($z > 3$) provide information about the growth of supermassive black holes and the evolution of active galactic nuclei (AGNs) in the early Universe, when it was at about 5–15% of its current age (An et al. 2020). The spatial density of blazars shows a peak at $z \sim 3$ (Diana et al. 2022), and even though the number of detections of distant quasars above this redshift keeps increasing, the fraction of distant radio-loud quasars is still limited (Bañados et al. 2021, Ighina et al. 2023, 2024). Given that the morphology and jet kinematics is known for even more limited number of distant blazars, single-dish radio continuum observations appear as an important approach to study them. Comprehensive analysis of the radio spectra of high-redshift blazars can help evaluate the physical processes applicable to them and determine the mechanisms that could power the central engine. The estimates of radio variability from multiwavelength long-term monitoring also serve as a valuable tool to investigate the quasar physical properties.

---

[†]Deceased
[*]Electronic address: lacerta999@gmail.com

The high-redshift ($z = 3.21$, Wilkes et al. 1983) source PKS 1614+051 is a bright radio-loud quasar which is classified as an FSRQ—a flat spectrum radio quasar—according to the BZCAT classification (Massaro et al. 2009). PKS 1614+051 is as well one of the first known peaked-spectrum (PS) radio sources, with a spectral peak (O'Dea 1990, O'Dea et al. 1991, Sotnikova et al. 2021) at 4–5 GHz, which makes it a high-frequency peaked radio source (HFP) (Dallacasa et al. 2000, Orienti et al. 2006a, Tinti et al. 2005). On the historical timescale its spectral shape remains relatively constant. A radio variability index of $\sim 0.10$ was found at 5 GHz on the historical timescale of about 43 yrs, and $\sim 0.20$ at 22 GHz in the RATAN-600 multifrequency monitoring program (Sotnikova et al. 2024). In general, the two leading mechanisms, synchrotron self-absorption (SSA, Pacholczyk 1970, Snellen et al. 1998) and free-free absorption (FFA, Bicknell et al. 1997), can be the cause of spectral turnover in the PS sources, producing the observed "peak frequency – size" anticorrelation (Fanti et al. 1990, O'Dea and Baum 1997). According to this relation, HFPs are very compact (few kpc) and therefore are considered either young or surrounded by a dense medium (O'Dea and Saikia 2021). The analysis of optical data (Husband et al. 2015, Orienti et al. 2010, Snellen et al. 2002, Stanghellini et al. 1993) showed the presence of companions around many HFPs hosted in galaxies, supporting the idea that the interactions between galaxies play a decisive role in the origin of the radio emission from young evolving objects.

PKS 1614+051 is interesting as a high-redshift HFP quasar which can be a good candidate to be a newly born radio source, with an age of $10^2$–$10^3$ yrs (Dallacasa et al. 2000). According to the "young scenario" (e.g. Blake 1970, Phillips and Mutel 1982), HFPs will develop into extended radio galaxies and quasars (e.g., FR I or FR II). As the peaked spectral shape is found in compact radio sources such as occasionally flaring blazars, when a new bright compact component is born, it is necessary to separate out the young intrinsically compact radio sources and the blazars. Thus, multifrequency and long-term measurements are needed to study the HFPs.

In this work, we present the results of a multiwavelength (MW) variability analysis of the blazar PKS 1614+051 on the long-term timescale of 27 yrs in the observer's frame of reference, covering the period of 1997–2024. We analyze the evolution of the PKS 1614+051 radio spectra at seven frequencies: 1, 2, 5, 8, 11, 22, and 37 GHz, using the RATAN-600 (SAO RAS) instantaneous spectra and the data from the RT-22 and RT-32 telescopes (CrAO RAS, IAA RAS). The optical studies were conducted with the SCORPIO-I spectrograph on the SAO RAS 6-meter (BTA) telescope and were aimed at obtaining long–slit spectra with a resolution $\sim 1$ nm across the entire optical spectral range at 380–740 nm. The $R$ band variability of the blazar was analyzed using the data taken with the SAO RAS 1-meter and 0.5-meter telescopes. Our goal is to understand the physical processes that govern the blazar variability and spectral evolution, using both radio and optical data across a wide frequency range. By examining the spectral characteristics, variability timescales, and absorption mechanisms, we aim to provide insights into the jet physics and the role of the surrounding environment in the evolution of high-redshift blazars.

The paper is organised as follows: details of the observations and the radio light curves are described in Section 2; the variability properties are presented in Section 3; in Section 4 we construct quasi-simultaneous radio spectra at 1–37 GHz with the aim to model them using the SSA and FFA mechanisms and to estimate the magnetic field temporary variation; in Section 5 we consider the results of the new optical spectral study of PKS 1614+051 with the BTA SCORPIO-I spectrograph. Finally, general re-sults of our work are discussed in Section 6. In this paper we used the standard flat $\Lambda$CDM cosmological model with $H_0 = 70$ km s$^{-1}$ Mpc$^{-1}$, $\Omega_{\rm m} = 0.3$, and $\Omega_{\rm vac} = 0.7$, in which the 1 mas angular scale corresponds to a projected linear scale of 7.5 pc at $z = 3.21$.

## 2. OBSERVED DATA

We have collected radio measurements on a long timescale of 27 yrs covering the frequency range of 1–37 GHz. The instruments and general details of the observations are given in Table 1. Since the observing frequencies of the RT-32 (5.05, 8.63 GHz) and RATAN-600 (4.7, 8.2 GHz) are close, we use their rounded values, 8 and 5 GHz, in the further analysis. The frequencies 21.7/22.3, 11.2, 2.3, and 0.96/1.2 GHz are also used in rounded form: 22, 11, 2, and 1 GHz.

The total historical radio spectrum of PKS 1614+051 was constructed based on the literature data taken from the CATS database (Verkhodanov et al. 2005, 1997). The blazar is a bright radio source and has been observed in almost 50 radio surveys in the frequency range of 0.076–37 GHz. The total number of literature data points is 913. In Table 2 we present a list of the main catalogs which have been used in this paper ($N_{\rm obs} \geq 5$).

### 2.1. Radio Continuum Observations

At the RATAN-600 radio telescope the observations have been made in 1997–2024 at 1–22 GHz. Part of the measurements have been published in Mingaliev et al. (2012), Sotnikova et al. (2021, 2019). The instantaneous RATAN-600 radio spectra were measured within 3–5 minutes at frequencies of 0.96/1.1/1.2, 2.3, 4.7, 7.7/8.2, 11.2, and 21.7/22.3 GHz (Kovalev et al. 1999, Parijskij 1993, Sotnikova 2020, Tsybulev 2011, Tsybulev et al. 2018, Udovitskiy et al. 2016, Verkhodanov 1997). The antenna and radiometer parameters are given in Table 3. The flux densities $S_\nu$, their errors $\sigma$, and average observing epochs are presented in Table 4, light curves at radio frequencies are shown in Fig. 1.

Additional measurements of PKS 1614+051 at 4.7 GHz have been obtained with daily cadence from 31.05.2019 to 07.06.2020 in the radio survey mode of RATAN-600. The observations were conducted with three/four-beam radiometer complex with a central frequency of 4.7 GHz (Table 3). The measurements from each of the 4.7 GHz subchannels were averaged. The dataset includes 337 observing epochs (Fig. 2). The averaged flux density is 1.1 Jy with a standard deviation of 0.03 Jy at 4.7 GHz. The average observing epochs (yyyy.mm.dd and yyyy.yy), flux densities, and their errors are presented in Table 5. Details of the observations and data processing are described in Kudryashova et al. (2024), Majorova et al. (2023).

The flux densities at frequencies of 5.05 and 8.63 GHz have been measured at several epochs from September 2022 to July 2024 using two RT-32 radio telescopes: Zelenchukskaya (Zc) and Badary (Bd) (Shuygina et al. 2019). All the antennas and receivers have similar parameters: bandwidth $\Delta f_0 = 900$ MHz for both receivers with the central frequencies specified above; beam width at the half-power level $HPBW = 7\rlap{.}'0$ and $3\rlap{.}'9$, respectively; flux density limit $\Delta F$ reaching about 20 mJy per scan with a time constant of 1 s for both frequencies under optimal observing conditions. The observations were performed in the elevation drift scan mode and processed with the original program package CV (Kharinov and Yablokova 2012) and the Database of Radiometric Observations. The observations and processing methods are described in detail in Vlasyuk et al. (2023). The flux densities $S_\nu$, their errors $\sigma$, and average observing epochs are presented in Table 4, the estimates of flux density at frequencies of 5 and 8 GHz are shown in Fig. 1.

The measurements at a frequency of 36.8 GHz (hereafter 37 GHz) have been obtained with the 22-meter RT-22 radio telescope. The beam-modulated receivers were used to acquire the data. The antenna temperature from the source was measured as a difference between the signals from the radiometer output in two antenna positions, when the radio telescope was pointed at the source alternately with one or the other receiving horns ("on-on observation" method). The observations of the blazar consisted of 5–20 such measurements to achieve the required signal-to-noise ratio. Details of the observations and data reduction are presented in Sotnikova et al. (2022), Volvach et al. (2023).

Table 1: The instruments used in the observations

| Telescope | Institution | Period | Passband |
|---|---|---|---|
| RATAN-600 | SAO RAS | 1997–2024 | 1–22 GHz |
| RT-32 | IAA RAS | 2022–2024 | 5, 8 GHz |
| RT-22 | CrAO RAS | 2005–2024 | 37 GHz |
| Zeiss-1000 | SAO RAS | 2023-2024 | optical $R$ |
| AS-500/2 | SAO RAS | 2023-2024 | optical $R$ |
| 48-inch Schmidt | ZTF Palomar | 2018–2023 | optical ZTF-$r$ |
| BTA | SAO RAS | 2024 | 350–750 nm |

Table 2: The literature data used in this paper: the instrument or catalog name from CATS (Col. 1), period of observations (Col. 2), number of measurements (Col. 3), frequency (Col. 4), and references (Col. 5)

| Catalog name | Period | $N_{obs}$ | Frequency, GHz | Reference |
|---|---|---|---|---|
| (1) | (2) | (3) | (4) | (5) |
| GBIMO | 1988–1994 | 578 | 2.25, 8.3 | Lazio et al. (2001) |
| CGR15 | 2008–2009 | 142 | 15 | Richards et al. (2011) |
| RCSP | 1980–1981 | 25 | 0.365–11.1 | Bursov et al. (1996) |
| GPSra | 2006–2010 | 24 | 1.1, 2.3, 4.8, 7.7, 11.2, 21.7 | Mingaliev et al. (2012) |
| MGPS2 | 1996–1997 | 22 | 2.3, 4.8, 7.7, 11.2, 21.7 | Murphy et al. (2007) |
| GLEAM | 2013–2014 | 21 | 0.076–0.220 | Hurley-Walker et al. (2017) |
| GPSDa | 1998–2000 | 11 | 1.4, 5.0, 8, 15, 22 | Dallacasa et al. (2000) |
| GPSTi | 2000 | 8 | 1.4, 1.7, 4.5, 5.0, 8.1, 8.5, 14.9, 22.5 | Tinti et al. (2005) |
| Kov97 | 1997 | 6 | 2.3, 3.9, 7.7, 11.2, 21.7 | Kovalev et al. (1999) |
| RATAN-600 | 2011–2016 | 5 | 1.2, 2.3, 4.8, 7.7, 11.2 | Sotnikova et al. (2019) |
| RATAN-600 | 2012–2020 | 10 | 2.3, 4.8, 7.7, 11.2, 22 | Sotnikova et al. (2021) |

The total period of observations with the RT-22 spans the time interval since May 2005 till June 2024. The flux densities, their errors, and average observing epochs are presented in Table 6. The light curve at frequency of 36.8 GHz is shown in Fig. 1. The data up to 2012 have previously been published in Vol'vach et al. (2015).

**Table 3**: The RATAN-600 continuum radiometer and antenna parameters for secondary mirrors 1 and 5: the central frequency $f_0$, bandwidth $\Delta f_0$, and detection limit for point sources per transit $\Delta F$. FWHM$_{\rm RA \times Dec.}$ is the angular resolution along RA and Dec., calculated for the average angles

| $f_0$, GHz | $\Delta f_0$, GHz | $\Delta F$, mJy beam$^{-1}$ | FWHM$_{\rm RA \times Dec.}$ |
|---|---|---|---|
| Secondary mirror 1 | | | |
| 22.3 | 2.5 | 50 | $0\rlap{.}''17 \times 1\rlap{.}'6$ |
| 11.2 | 1.4 | 15 | $0\rlap{.}''34 \times 3\rlap{.}'2$ |
| 8.2 | 1.0 | 10 | $0\rlap{.}''47 \times 4\rlap{.}'4$ |
| 4.7 | 0.6 | 8 | $0\rlap{.}''81 \times 7\rlap{.}'6$ |
| 2.25 | 0.08 | 40 | $1\rlap{.}'7 \times 16'$ |
| 1.25 | 0.08 | 200 | $3\rlap{.}'1 \times 27'$ |
| Secondary mirror 5 | | | |
| $4.40 - 4.55$ | 0.15 | 10 | $1\rlap{.}'5 \times 35'$ |
| $4.55 - 4.70$ | 0.15 | 10 | $1\rlap{.}'5 \times 35'$ |
| $4.70 - 4.85$ | 0.15 | 10 | $1\rlap{.}'5 \times 35'$ |
| $4.85 - 5.00$ | 0.15 | 10 | $1\rlap{.}'5 \times 35'$ |

### 2.2. Optical observations

The spectroscopy of PKS 1614+051 was performed on the night of 14/15 July 2024 of director's discrete time at the 6-meter telescope (BTA) with the SCORPIO-I spectrograph (Afanasiev and Moiseev 2005). The atmospheric conditions were quite good: transparency was excellent and the seeing (measured as the full width at half maximum of stellar spectra in individual exposures) was about $1\rlap{.}''8$ even for an air mass of about 2.0. We used the $1\rlap{.}''2$–wide slit in combination with the VPH grating 550G from the SCORPIO-I standard set.[1]

---

[1] https://www.sao.ru/hq/lsfvo/devices/scorpio/scorpio.html

This instrumental setup along with a $2048 \times 2048$ px back-illuminated E2V chip CCD 42-40 as a detector provide a spectral resolution of about 1 nm across the total spectral range between 380 and 740 nm. The total exposure was combined from three individual 10-min exposures. The spectra were processed with the software package described in Vlasyuk (1993). The shape of the spectra was corrected using the observations of spectrophotometric standards from Massey et al. (1988).

The photometric data (February 2023 – September 2024) have been taken with the 1-metre Zeiss-1000 and 0.5-metre AS-500/2 optical reflectors. The additional data within the time interval since 2018 were taken from the data archives of the ZTF project (Bellm et al. 2019). Details of the Zeiss-1000 instrumental setup are described in Komarov et al. (2020), Vlasyuk et al. (2023). The main characteristics of the instrumental complex of an AS-500/2 reflector are described in Valyavin et al. (2022). To improve the efficiency of the studies with the 0.5-metre telescope, a back-illuminated electron-multiplying CCD camera Andor IXon$^{\rm EM}$+897 was installed in the Cassegrain focus in July 2023.

This $512 \times 512$ px CCD camera with a quantum efficiency of $\sim 90\%$ in the 450–700 nm range provides a $7'$ field of view with a $0\rlap{.}''82$/px data sampling (1 px = 16 $\mu$m). The system readout noise is about 6 $e^-$, the CCD operating temperature has been chosen to minimize the dark current. Thus, the statistical dark-sky background noise in broadband photometric studies dominates over the system noise at exposures longer then 30 s. Both CCD photometers are equipped with similar sets of filters, which are close to the standard broadband Johnson–Cousins system, given the sensitivity of both CCDs. The typical exposure time for observations of PKS 1614+051 was 300 s for Zeiss-1000 and 90–120 s for the AS-500/2.

To obtain blazar flux estimates, we performed the standard reduction steps, which were described in Bychkova et al. (2018), Vlasyuk (1993). We collected the optical data as the instrumental magnitudes of the source and secondary standard stars in the same field. The latter were obtained by ourselves from flux estimates of the objects in the PKS 1614+051 field and the standard stars around the blazars S5 0716+71 and Ton 599 from González-Pérez et al. (2001).

Table 4: Flux density measurements with RATAN-600 and RT-32: observing date in yyyy.mm.dd (Col. 1), observing date in yyyy.yy (Col. 2), flux densities at frequencies from 1 to 22 GHz and their uncertainties in Jy (Cols. 3–8), and the instrument (Col. 9). The measurements published in Mingaliev et al. (2012), Sotnikova et al. (2021, 2019) are marked with an asterisk in the database. A short fragment is shown; the full version is available online in VizieR database

| Date | Epoch | $S_{22} \pm \sigma$, Jy | $S_{11} \pm \sigma$, Jy | $S_8 \pm \sigma$, Jy | $S_5 \pm \sigma$, Jy | $S_2 \pm \sigma$, Jy | $S_1 \pm \sigma$, Jy | Instrument |
|---|---|---|---|---|---|---|---|---|
| (1) | (2) | (3) | (4) | (5) | (6) | (7) | (8) | (9) |
| 1997.03.18 | 1997.21 |  | 0.7 ±0.01 |  | 0.81±0.01 | 0.69±0.02 | 0.2 ±0.02 | RATAN-600 |
| 1997.06.21 | 1997.47 | 0.32±0.04 | 0.71±0.01 | 0.84±0.05 | 0.91±0.04 | 0.77±0.22 |  | RATAN-600 |
| 1997.09.13 | 1997.70 |  | 0.59±0.12 | 0.82±0.01 | 0.76±0.03 | 0.78±0.03 |  | RATAN-600 |
| 1997.12.10 | 1997.94 | 0.33±0.06 | 0.68±0.03 | 0.78±0.03 | 0.86±0.01 | 0.55±0.01 |  | RATAN-600 |
| 1998.04.16 | 1998.29 | 0.27±0.08 | 0.59±0.03 | 0.8 ±0.02 | 0.97±0.02 | 0.74±0.03 | 0.16±0.03 | RATAN-600 |

Table 5: The RATAN-600 daily measurements of the blazar PKS 1614+051 flux densities in 2019–2020: date and epoch of observations, in yyyy.mm.dd and yyyy.yy format, respectively (Cols. 1–2), flux densities at 5 GHz and their errors (Col. 3). A short fragment is shown; the full version is available online in VizieR database

| Date | Epoch | $S_5 \pm \sigma$, Jy |
|---|---|---|
| (1) | (2) | (3) |
| 2019.05.31 | 2019.41 | 1.16 ± 0.01 |
| 2019.06.01 | 2019.42 | 1.18 ± 0.08 |
| 2019.06.02 | 2019.42 | 1.17 ± 0.01 |
| 2019.06.03 | 2019.42 | 1.15 ± 0.01 |
| 2019.06.04 | 2019.42 | 1.15 ± 0.01 |
| 2019.06.05 | 2019.43 | 1.13 ± 0.01 |

Table 6: The blazar PKS 1614+051 flux density measurements at 37 GHz obtained with RT-22: date and epoch of observations, in yyyy.mm.dd and yyyy.yy format, respectively (Cols. 1–2), flux densities at 37 GHz and their errors (Col. 3). A short fragment is shown; the full version is available online in VizieR database

| Date | Epoch | $S_{37} \pm \sigma$, Jy |
|---|---|---|
| (1) | (2) | (3) |
| 2005.05.29 | 2005.41 | 0.18±0.06 |
| 2005.06.22 | 2005.47 | 0.22±0.08 |
| 2005.07.05 | 2005.51 | 0.21±0.05 |
| 2005.08.04 | 2005.59 | 0.28±0.07 |
| 2005.08.19 | 2005.63 | 0.25±0.08 |
| 2005.10.17 | 2005.79 | 0.19±0.07 |

In order to provide a joint analysis of the optical and radio data, we averaged the measurements over individual nights and transformed the resulting values into fluxes according to the constant from Mead et al. (1990). The optical data are collected from 379 observing nights spanning between March 2018 and September 2024. The light curve combined from our original results and the ZTF data is presented in Fig. 3.

### 2.3. Multi-wavelength Light Curves

Figure 1 shows the MW light curves of PKS 1614+051 collected from March 1997 until July 2024 at 1–22 GHz, and from 2005 May to July 2024 at 37 GHz. The low-frequency light curves at 1–2 GHz contains extended gaps due to strong radio interference, the light curve at 22 GHz is also has multiple gaps due to worse observing conditions. The light curves show slow flux density variations with one prominent flare, which is detected well at 5–22 GHz and started at the end of 2009. At frequencies of 8–22 GHz this flare had finished approximately in the middle of 2022, at 5 GHz the flare has not finished yet. For 37 GHz in Fig. 1 we can see a slightly other behavior: there was one prominent flare with a maximum flux density $S_{37} = 0.55$ Jy in 2014 Jan and several faint but well localized flares in 2009, 2020, and 2022.

## 3. VARIABILITY ANALYSIS

In this section we analyze the variability properties using common approaches such as calculation of the variability $V_S$ and modulation $M$ indices, fractional variability $F_{\rm var}$, the structure functions (SF) method, and the discrete correlation functions (DCFs) in order to quantify the normalized difference between the maximum and minimum flux density, the time scale of variations, the correlation and time lags between flares at different frequencies.

### 3.1. Variability Indices

First of all, we note that the radio light curves demonstrate one great flare only. This flare is well tracked at 37 GHz and 11 GHz, while is not obvious at 22 GHz because of sparse data. The remarkable feature of this flare is its long time scale: about 5 years at 37 GHz, approximately 10–15 years at 11 GHz and 8 GHz, while at 5 GHz this flare is strongly smeared and the radio light curve has not reached its preflaring level yet.

The variability index $V_S$ is calculated as in Aller et al. (1992) from the maximum $S_{\rm max}$ and minimum $S_{\rm min}$ flux densities within the period of observations:

$$V_S = \frac{(S_{\rm max} - \sigma_{S_{\rm max}}) - (S_{\rm min} + \sigma_{S_{\rm min}})}{(S_{\rm max} - \sigma_{S_{\rm max}}) + (S_{\rm min} + \sigma_{S_{\rm min}})}, \quad (1)$$

where $\sigma_{S_{\rm max}}$ and $\sigma_{S_{\rm min}}$ are measurements uncertainties. The uncertainty $\Delta V_S$ of the variability index was calculated as

$$\Delta V_S = \frac{2S_{\rm min}(\sigma_{S_{\rm min}} + \sigma_{S_{\rm max}})}{(S_{\rm min} + S_{\rm max})^2}. \quad (2)$$

Because we have one great flare for PKS 1614+051 in the radio light curves, the variability index can be considered as its relative amplitude. Since $V_S$ is determined by the two extreme flux density values only, it is useful to characterise the light curve variability on the whole, taking into account all flux density measurements. We used fractional variability $F_{\rm var}$ (Vaughan et al. 2003) to quantify the variability level relative to its mean value at each frequency:

$$F_{\rm var} = \sqrt{\frac{V^2 - \bar{\sigma}_{\rm err}^2}{\bar{S}^2}}, \quad (3)$$

where $V^2$ is variance, $\bar{S}$ is the mean flux density, and $\bar{\sigma}_{\rm err}$ is the root mean square error. The uncertainty of $F_{\rm var}$ is determined as:

$$\triangle F_{\rm var} = \sqrt{\left(\sqrt{\frac{1}{2N}}\frac{\bar{\sigma}_{\rm err}^2}{F_{\rm var} * \bar{S}^2}\right)^2 + \left(\sqrt{\frac{\bar{\sigma}_{\rm err}^2}{N}}\frac{1}{\bar{S}}\right)^2}. \quad (4)$$

However, we cannot use $F_{\rm var}$ in the case of strong noisy data, therefore we failed to estimate $F_{\rm var}$ for PKS 1614+051 at 37 GHz. Nevertheless, we can estimate the variability level relative to the mean flux density $\bar{S}$ without taking into account measurements uncertainties by using the modulation index $M$ as in Kraus et al. (2003):

$$M = \frac{\sigma_S}{\bar{S}}, \quad (5)$$

where $\sigma_S$ is the flux density standard deviation.

The calculated $V_S$, $F_{\rm var}$, and $M$ are presented in Table 7. We note that the variability level does not show a prominent frequency dependence and lies in the range of 15–30%. The $F_{\mathrm{var}}$ values lie in the range of $\sim$0.2–0.3, and $M$ is in the range of 0.1–0.3. The optical indices $F_{\mathrm{var}}$ and $M$ are also quite low, on the time scale of $\sim 6$ yrs they are 0.10 and 0.07, respectively. It is well known that refractive interstellar scintillations (RISS) can cause variability of extragalactic radio sources (e.g., Ross et al. 2022, Walker 1998). We have estimated the potential RISS contribution into the measured variability level at the observing frequencies from 1 GHz to 37 GHz. The transition frequency equals to 3.42 GHz for the coordinates of PKS 1614+051 according to the RISS19[2] code (Hancock et al. 2019); thus, we have a strong scattering mode for 1 GHz and 2 GHz and a weak one for the other frequencies. For a point source, the ratio between the observing and transition frequencies determines the flux density modulation level $m$ and its timescale $t$ in these two scattering modes according to formulas from Walker (1998). If taken into account, the source's size $\theta_s$ decreases $m$ and increases $t$. We aimed to estimate the maximum possible RISS contribution into the measured variability level, therefore we chose the minimum size for a stationary synchrotron radio source (Kellermann and Owen 1988) as a proxy for the source's size: $\theta_s = \theta_{\min} = 0.6 \times \sqrt{S}/\nu$, where $S$ is the flux density at an observing frequency $\nu$. We used the median flux density during the observing period as $S$. We took a set of source's sizes $\theta_{\min}$, $2\times\theta_{\min}$, $4\times\theta_{\min}$, $8\times\theta_{\min}$, and $16\times\theta_{\min}$ in order to illustrate the dependence $m = m(\nu, \theta_s)$. The results of calculations are shown in Fig. 4, where we have also plotted the modulation index according to flux density measurements. We see that the RISS contribution can be significant only at 1 GHz and 2 GHz if the source size $\theta_s = (1$–$2)\theta_{\min}$, we can neglect the RISS effects at higher frequencies and with a larger source's size.

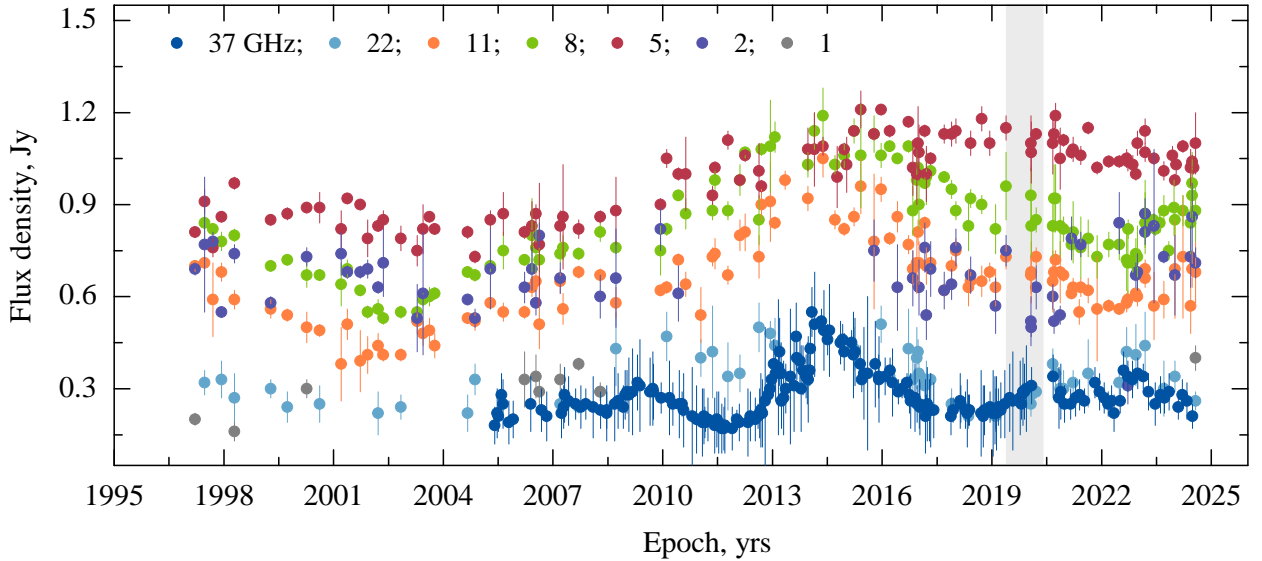

**Figure 1**: The MW light curves of PKS 1614+051 in 1997–2024. The grey area is the time interval of the daily observations at 5 GHz (see the description of RATAN-600 observations).

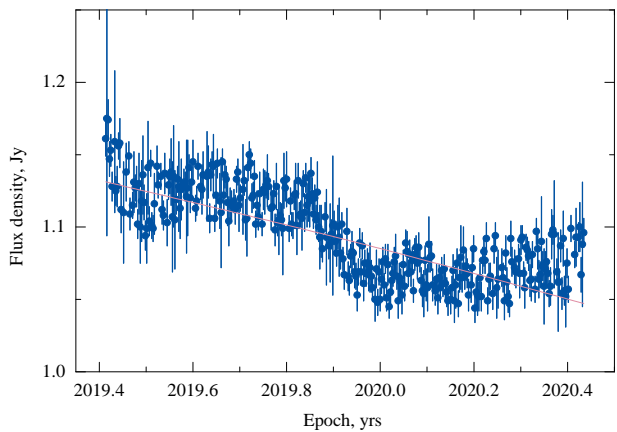

**Figure 2**: The light curve of PKS 1614+051 in 2019–2020 at 5 GHz, measured with RATAN-600. The red line shows the flux density changing with a negative slope of -0.08 Jy yrs$^{-1}$.

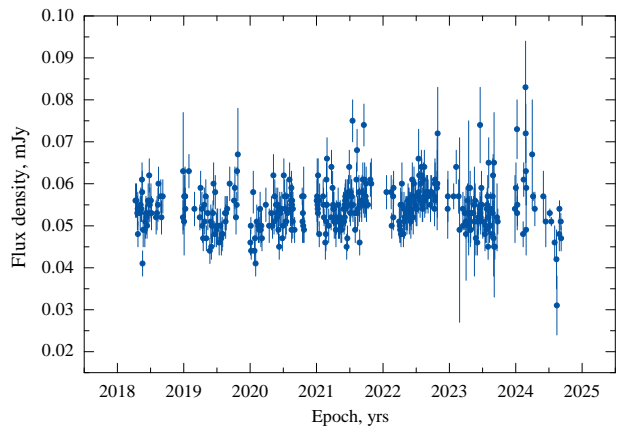

**Figure 3**: Optical $R$-band light curve of PKS 1614+051 in 2018–2024 according to our data and estimates from ZTF data archive.

---

[2] https://github.com/PaulHancock/RISS19

**Table 7**: The variability indices of PKS 1614+051

| Band, GHz | $V_S$ | $\Delta V_S$ | $M$ | $F_{\mathrm{var}}$ | $\Delta F_{\mathrm{var}}$ |
|---|---|---|---|---|---|
| 37 | 0.31 | 0.11 | 0.27 | – | – |
| 22 | 0.29 | 0.08 | 0.23 | 0.12 | 0.04 |
| 11 | 0.33 | 0.07 | 0.20 | 0.18 | 0.01 |
| 8 | 0.33 | 0.04 | 0.18 | 0.16 | 0.07 |
| 5 | 0.21 | 0.03 | 0.12 | 0.10 | 0.07 |
| 2 | 0.38 | 0.05 | 0.16 | 0.09 | 0.02 |
| 1 | 0.31 | 0.07 | 0.23 | 0.17 | 0.06 |
| optical $R$ | 0.31 | 0.07 | 0.10 | 0.06 | 0.07 |

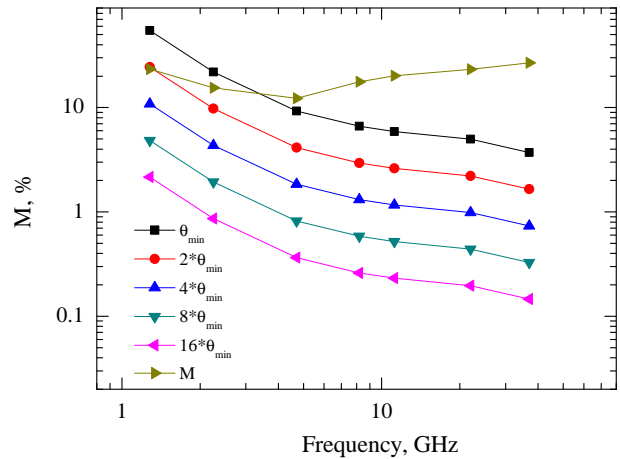

**Figure 4**: The potential RISS contribution as a function of the source's size and observed frequency. The measured modulation index $M$ is shown by brown triangles.

### 3.2. Structure Function Analysis

The structure function (SF) is a widely used method to identify typical timescales in non-stationary processes, providing quantitative measures of variability and insights into the mechanisms driving these variations (Heidt and Wagner 1996). In this study, we employed the first-order SF normalized to the variance of the signal $\sigma^2$:

$$D_1(\tau) = \langle\{[f(t) - f(t+\tau)]\}^2\rangle, \qquad (6)$$

where $f(t)$ is the signal at a time $t$, and $\tau$ is the lag.

We follow the method described in Hughes et al. (1992), Simonetti et al. (1985). To construct intervals for $k = 1, 2, ..., L$, the initial interval $k = 1$ was selected to be comparable to the average time between observations, while ignoring significant observational gaps. It was 27–33 days, and the final interval was determined based on the total time span of the observations.

We estimated the uncertainties using the bootstrap method. For each frequency, a model light curve was generated by applying a 2-day boxcar smoothing, followed by subtracting the smoothed curve from the original data to obtain the residuals. These residuals were then added randomly to the smoothed curve, simulating new light curves. This procedure was repeated 1000 times to recompute the SF, and the confidence intervals were defined by the point where only 5 of the simulated data points deviated, providing an estimate at the 99 per cent confidence level.

Due to the highly uneven time series at some frequencies, we constructed the SF in two ways: using the initial light curves (blue points in Fig. 5) and the interpolated ones (pink points). We used cubic Hermite interpolation to get the interpolated light curves. In most cases, $\tau$ is systematically higher for the interpolated light curves (Table 8). At 5 GHz, where the flare is not finished yet, the initial light curve does not allow calculation of the variability timescale, and the interpolated one gives $\tau_{\rm obs} \simeq 15.4$ years.

The analysis reveals different variability timescales at 8–37 GHz (Table 8), varying from 1.7 to 6.9 yrs in the observer's frame of reference, which corresponds to $\tau_{\rm rest} = 0.2$–$0.8$ yrs accepting the Doppler factor $\delta = 2$ (see Section 4.4.4) or $\tau_{\rm rest} = 0.3$–$1.3$ yrs for $\delta = 1.27$ (Liodakis et al. 2018); $\tau_{\rm rest} = \tau_{\rm obs}/(1+z) \cdot \delta$. The obtained values are comparable with the variability scale of blazars at the intermediate and low redshifts.

While SF analysis must always be approached with caution (Emmanoulopoulos et al. 2010) due to potential artifacts caused by data gaps and limitations in time series length, the results for PKS 1614+051 at higher frequencies (8–37 GHz) appear to capture genuine variability trends. The lack of characteristic structures at lower frequencies (2 and 5 GHz) likely reflects either longer variability timescales beyond the observation period or insufficient sampling.

### 3.3. Correlations Between the Light Curves at Different Frequencies

The discrete correlation functions (DCFs, Edelson and Krolik 1988) have been calculated between the 5, 8, 11, 22, and 37 GHz flux density curves of PKS 1614+051 during 1997–2024. The DCF method was applied in the same way as in Vlasyuk et al. (2023) and using the Python-based software of Robertson et al. (2015). We varied the bin widths DT = 30, 60, and 90 days for different pairs of frequencies to obtain the most significant results. For estimation of the confidence levels we utilized Monte-Carlo simulations which are outlined in (Emmanoulopoulos et al. 2013) with the description of correspondence software.

**Table 8**: The variability timescales of the SF in the observer and rest-frame ($\tau_{\rm obs}$, $\tau_{\rm rest}$, respectively) at 2, 5, 8, 11, 22, and 37 GHz, calculated for PKS 1614+051 without ($\tau_1$) and with interpolation ($\tau_2$), respectively

| Band, GHz | $\tau_{obs,1}$, yrs | $\tau_{rest,1}$, yrs | $\tau_{obs,2}$, yrs | $\tau_{rest,2}$, yrs |
|---|---|---|---|---|
| 2 | – | – | – | – |
| 5 | – | – | 15.4 | 1.8 |
| 8 | 6.9 | 0.8 | 8.7 | 1.0 |
| 11 | 4.3 | 0.5 | 8.7 | 1.0 |
| 22 | 4.3 | 0.5 | 5.5 | 0.7 |
| 37 | 1.7 | 0.2 | 3.1 | 0.4 |

The DCFs between the pairs of frequencies over the entire period with confidence levels $\geq 2\sigma$ as well as corresponding time lags are presented in Table 9 and Fig. 6. The DCF analysis shows large time delays up to several years in the observer's frame of reference, which corresponds to 1–1.5 years in the rest frame. Minimum delays of 0.6 and 0.8 years are observed for the pairs "11 GHz – 8 GHz" and "37 GHz – 22 GHz." The time lags between the light curve at 22 GHz and the 11, 8, and 5 GHz light curves are from 3.4 to 6.4 years, and reach almost 4.4 years between 8 GHz and 5 GHz. The result reflects the scenario where the time lags at lower frequencies are due to greater optical opacity of the matter and greater synchrotron self-absorption. The only exception from this picture is the "37 GHz – 11 GHz" pair of frequencies (see Fig. 6), where the DCF has its maximum at a negative time lag (i.e., the light curve at the higher frequency looks lagging that at the lower frequency). Taking into account the overall behaviour of the rest of the light curves and the $\sim 2\sigma$ significance of this peak, we tend to consider this particular DCF maximum accidental.

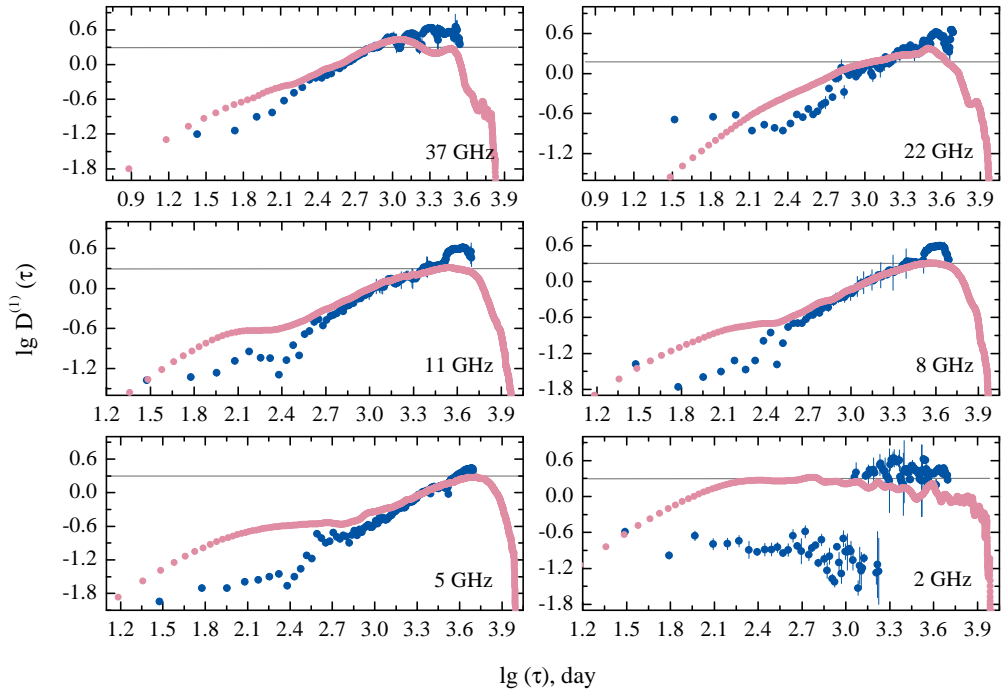

**Figure 5**: The SFs for the total flux variations at 2, 5, 8, 11, 22, and 37 GHz obtained without (dark blue) and with interpolation (pink).

**Table 9**: Time lags between the light curves at different frequencies

| Bands, GHz | lag, yrs | DT, days | DCF peak value and its significance level |
|---|---|---|---|
| 37 vs 22 | 0.8 | 30 | $0.64 \pm 0.28$ ($2\sigma$) |
| 37 vs 5  | 1.2 | 60 | $0.38 \pm 0.11$ ($2\sigma$) |
| 22 vs 11 | 3.4 | 90 | $0.68 \pm 0.14$ ($2\sigma$) |
| 22 vs 8  | 2.9 | 90 | $0.73 \pm 0.17$ ($2\sigma$) |
| 22 vs 5  | 6.4 | 90 | $0.61 \pm 0.18$ ($2\sigma$) |
| 11 vs 8  | 0.6 | 60 | $0.86 \pm 0.16$ ($2\sigma$) |
| 11 vs 5  | 2.7 | 60 | $0.81 \pm 0.19$ ($3\sigma$) |
| 8 vs 5   | 4.4 | 60 | $0.84 \pm 0.19$ ($3\sigma$) |

### 3.4. Short-Term Radio Variability

The daily observations at 5 GHz carried out from May 2019 to June 2020 are presented in Fig. 2. The light curve shows slow decrease during the year of observations. The maximum flux density is $1.18 \pm 0.08$ Jy, the minimum is $1.04 \pm 0.01$ Jy.

The 5 GHz radio variability is quite low: $V_S = 0.02 \pm 0.08$, $F_{\rm var} = 0.02 \pm 0.001$, $M = 0.03$. These small values mean that the RISS contribution can be significant in the revealed flux density variability on short timescales. We have $\theta_{\min} = 0.13$ mas for the median flux density. The modulation level of flux density $m$ is about 2% for the source's size $\theta_s = 4 \times \theta_{\min} = 0.52$ mas implying RISS dominance, and is about 1% for $\theta_s = 8 \times \theta_{\min} = 1.04$ mas in the case of equal RISS contribution compared to the intrinsic variability. We note that the RISS timescale $t$ is 1.5–3 days under the above assumptions, thus the RISS effects cannot be smoothed during one measurement with RATAN-600 (few minutes). The source's size $\theta_s = 0.5$–$1.0$ mas at 5 GHz is expected for PKS 1614+051 based on VLBI observations (Koryukova et al. 2022, Pushkarev and Kovalev 2015). We conclude that the RISS effects can be significant in our daily measurements of PKS 1614+051.

The first-order 5 GHz structure function was calculated for two different cases. The first one was calculated for the initial data (top panel in Fig. 7), and the second was applied to the light curve after subtracting the linear trend (bottom panel in Fig. 7), which was subtracted to exclude the long-term variability trend. The initial time lag $\tau$ for both cases was 1 day. The SF for the initial light curve does not reach the second plateau, reflecting that variability timescale is greater than the timescale of the PKS 1614+051 monitoring (Simonetti et al. 1985). The SF shows a plateau at $\tau_{obs} \sim 100$ days (or 25 days in the rest frame) for the light curves after the exclusion of the long-term linear trend. According to Hughes et al. (1992), an SF slope $b < 1$ is between the flicker noise and the shot or random-walk noise (the process is prolonged during arbitrary times).

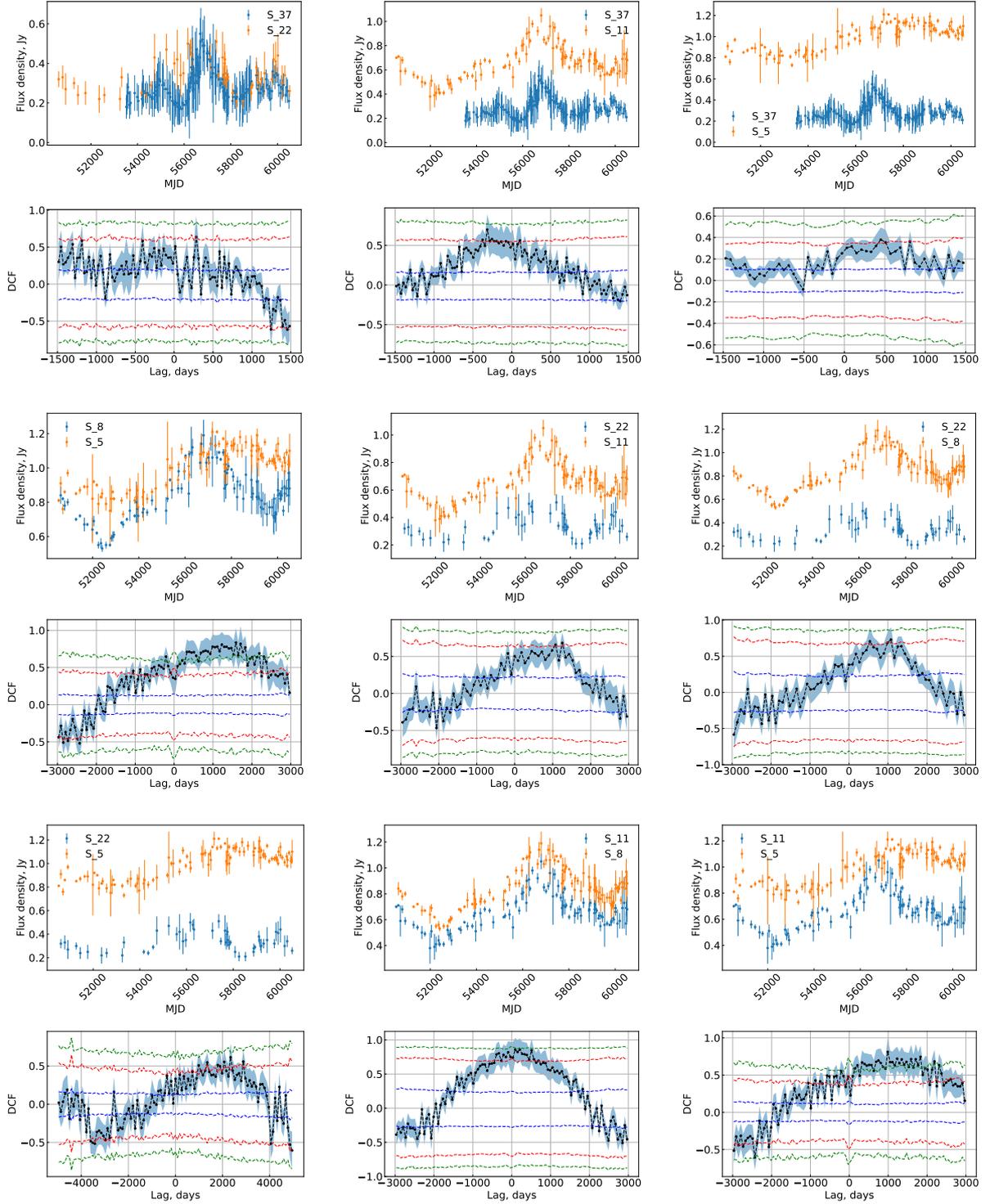

**Figure 6**: The DCFs for different combinations of frequency bands. Lags are given in days. The blue, red, and green lines represent the $1\sigma$, $2\sigma$, and $3\sigma$ significance levels, respectively. The blue areas correspond to the DCF uncertainties. Positive lags correspond to the situation where the emission at a lower frequency (orange measurements) lags the emission at a higher frequency (blue measurements).

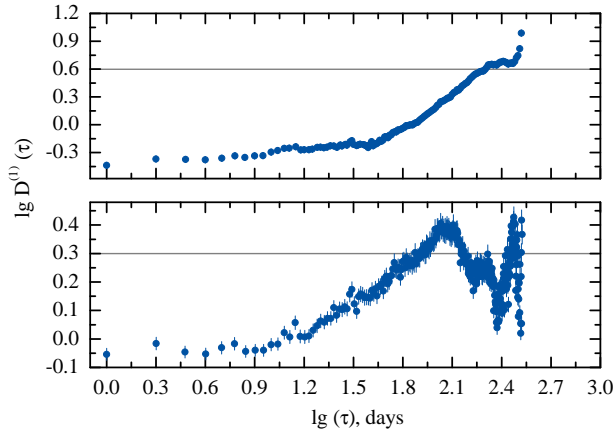

**Figure 7**: The structure function of the daily measurements in 2019–2020 at 5 GHz: the upper panel shows the SF for the initial light curve, in the bottom panel there is the SF calculated for the light curves with the linear trend subtracted.

To search for possible short-time periodicity, the Lomb–Scargle (L–S) periodogram (Lomb 1976, Scargle 1982) was calculated. The calculations were performed with the Python module `GLS` developed by Zechmeister and Kürster (2009). The L–S periodogram at 5 GHz does not show any significant peaks with FAP $\leq 1\%$ both for the initial dataset and for the data after subtracting the linear trend.

## 4. QUASI-SIMULTANEOUS RADIO SPECTRA MODELING

### 4.1. Quasi-Simultaneous Spectra

Using the measurements from several telescopes we constructed quasi-simultaneous spectra which differ from the RATAN-600 instantaneous spectra in that the flux densities are averaged within a certain time period. This allows the data obtained on different dates to be combined in the analysis of the spectrum shape. The time period should be short enough, typically of the order of 30 days, to consider the spectrum to be stable within this interval.

We took a time interval of 39 days so that to get the maximum number of quasi-simultaneous spectra with at least four frequencies with flux densities measured and possibly a measurement at 37 GHz. To analyze the two-slope shape of the spectrum according to the synchrotron self-absorption (SSA spectrum equation from Türler et al. (2000)) and inhomogeneous free-free absorption (FFA, Bicknell et al. 1997) models, we further selected only those spectra where the measurements had been available both for frequencies lower and greater than 5 GHz. This gave us a total of 48 spectra where the models could be applied.

In Fig. 8 we present the total radio spectrum of PKS 1614+051, including the literature data (grey points). The continuum spectra covers a frequency range of 0.07–40 GHz and can be described by a sum of two major contributions. One of them is the high-frequency component (HFC, Kovalev et al. 2002) that dominates at frequencies of 3–40 GHz and is associated with a parsec-scale jet. The HFC properties are determined by the variable jet spectra with a maximum at about 4–5 GHz. The low-frequency component (LFC) corresponds to the optically thin synchrotron emission of extended, up to kiloparsec scales, structures dominating at frequencies lower 1 GHz and having a peak frequency lower than 0.1 GHz due to their characteristic size (Slish 1963).

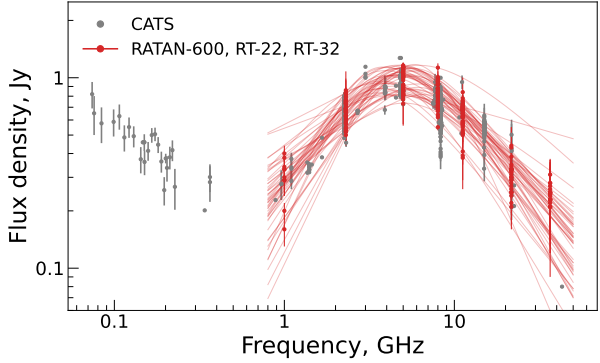

**Figure 8**: The broadband radio spectra of PKS 1614+051. The data from RATAN-600, RT-32, and RT-22 are obtained in 1997–2024 (red points connected by the SSA model lines). The measurements from the CATS database are shown with grey points.

### 4.2. SSA and FFA Models

The obtained quasi-simultaneous spectra could be described by several absorption models (see, e.g., Shao et al. 2022). In this paper we consider two of them: synchrotron self-absorption (SSA) and inhomogeneous free–free absorption (FFA).

The SSA radio spectrum is characterized by a peaked shape. It is commonly believed that the absorption is caused by the high density of emitting electrons. At the frequencies where the emission is self-absorbed, the spectrum rises steeply, and at the higher frequencies, where the emission is optically thin, the spectrum falls. At the frequency $\nu_m$ where the spectral index $\alpha$ (we assume $S \sim \nu^\alpha$) changes from positive to negative, the optical depth $\tau_m$ is approximately unity. Assuming a homogeneous self-absorbed incoherent synchrotron radio source with a power-law electron energy distribution with the power-law index $\gamma$ ($\alpha = -(\gamma-1)/2$), the SSA spectrum can be fitted as in Pacholczyk (1970), Tingay and de Kool (2003), Türler et al. (1999), van der Laan (1966):

$$S_\nu = S_m \left(\frac{\nu}{\nu_m}\right)^{\alpha_{\text{thick}}}$$
$$\times \frac{1 - \exp\left(-\tau_m \left(\frac{\nu}{\nu_m}\right)^{\alpha - \alpha_{\text{thick}}}\right)}{1 - \exp(-\tau_m)}, \quad (7)$$

where the optical depth

$$\tau_m \sim \frac{3}{2}\left(\sqrt{1 - \frac{8\alpha}{3\alpha_{\text{thick}}}} - 1\right). \quad (8)$$

The spectral index $\alpha = \alpha_{\text{thin}}$ provides information about the energy distribution of emitting particles, the spectral index $\alpha_{\text{thick}}$ describes the optically thick emission part at frequencies lower than $\nu_m$, at which flux density $S_m$ reaches its maximum. The four free parameters $S_m$, $\nu_m$, $\alpha_{\text{thick}}$, and $\alpha_{\text{thin}}$ are derived from radio spectra modeling.

The FFA processes are an alternative way to explain the peaked shape of radio spectra. The interaction of the propagating jet with the ambient interstellar medium can form an ionized screen external to the radio source (Bicknell et al. 1997). Thus, the FFA processes can play an important role under the conditions of high density of surrounding gas. It was shown that inhomogeneous FFA models can successfully describe broadband radio spectra of extragalactic sources (e.g., Ross et al. 2022, Shao et al. 2022):

$$S_\nu = S_{\text{norm}}(p+1)\left(\frac{\nu}{\nu_p}\right)^{2.1(p+1)+\alpha}$$
$$\times \gamma\left[p+1, \left(\frac{\nu}{\nu_p}\right)^{-2.1}\right], \quad (9)$$

where $S_{\text{norm}}$, $\nu_p$, and $\alpha$ are the normalization parameter, the turnover frequency, and the spectral index. The parameter $p$ describes the distribution of absorbing clouds, and $\gamma\left[p+1, \left(\frac{\nu}{\nu_{\text{peak}}}\right)^{-2.1}\right]$ is the incomplete gamma function of order $p+1$.

### 4.3. Model Fitting with Bayesian Approach

Although the spectrum parameters could simply be obtained from non-linear least squares fitting, this would not allow us to derive their uncertainties. More reliable parameter values along with the uncertainties can be derived from Bayesian statistics.

According to Bayes' theorem, the *posterior* probability that our model, e.g. the SSA formula, has the best-fit parameters $\theta$, given the observed data $D$, is

$$\ln P(\theta|D) \propto \ln P(D|\theta) + \ln P(\theta), \quad (10)$$

where $P(D|\theta)$ is the *likelihood* of the data given $\theta$, and $P(\theta)$ is the probability of $\theta$ itself. Although the exact value of the latter is usually unknown, as well as the coefficient of the proportion in equation 10, we can assume some *prior* probability distribution for model parameters $\theta$ and thus obtain the posterior probability distribution $P(\theta|D)$ that would give us the means to estimate the best-fit parameters and their uncertainties. For instance, the priors could be rather loose uniform distributions of the SSA or FFA parameters within their possible ranges, e.g., $0 < \nu_{\text{peak}} < 20$ GHz for the peak frequency.

Assuming that the deviations of the data from the model are normally distributed, the probability of a deviation is described by a Gaussian, the likelihood is the product of these probabilities, and the log-likelihood is

$$\ln P(D|\theta) = -\frac{1}{2} \sum \left( \frac{(y_i - \hat{y}_i)^2}{\sigma_i^2} + \ln(2\pi\sigma_i^2) \right), \quad (11)$$

where $y_i$ are the measurements, $\hat{y}_i$ are the corresponding values predicted by the model, and $\sigma_i$ are the measurement uncertainties. The second term under the sum can be omitted in further optimization.

As mentioned above, an analytical solution for the expected value of $\theta$ is still intractable from equation (10), as this requires calculation of a complex multidimensional integral. The solution is the numerical Markov chain Monte Carlo (MCMC) method (e.g., Goodman and Weare 2010) that allows one to sample the posterior probability distribution having the prior, the likelihood, and the data. The method maps the posterior with a number of "walkers," the steps of which are written in chains of parameters $\theta$, and the probability of a particular parameter value is distributed according to the posterior. We used the MCMC implementation from the `emcee`[3] library (Foreman-Mackey et al. 2013) along with the `corner`[4] library (Foreman-Mackey 2016) for visualization.

Having the posterior represented by a sample of parameter values saved in the chains, one can evaluate the best-fit parameters by various ways. We preferred the approach where the best fit are the medians of the marginal posterior distributions, and the uncertanties are the 16th and 84th percentiles of the sample. These are reasonable estimates for asymmetric posterior distributions, which are often the case. We did not use the parameters corresponding to the maximum likelihood, as this approach does not take into account the entire posterior distribution and is strongly dependent on the exact positions of the observed datapoints, which may be subject to unaccounted random factors. The maxima of marginal posterior distribution also cannot be used because they do not take into account correlations between the parameters, resulting in an unsatisfactory fit of the observed spectrum in the case of asymmetric posteriors. In Fig. 9 we show an example of Bayesian fitting.

Using the technique described, we modeled the quasi-simultaneous spectra using both models: SSA and inhomogeneous FFA. Interestingly, in the case of the latter, the medians of the marginal posteriors fit the observed spectrum unsatisfactorily. In this case, the only choice is taking the maximum likelihood parameters as the best fit (which is almost identical to least squares), while estimating the errors from Bayesian fitting.

For further modeling we compared how the models fit the observed data by calculating the Bayes factor from marginal likelihoods (evidence) using the UltraNest library[5] (Buchner 2021) and similar priors for the two models. While the Bayes factor varies from spectrum to spectrum, its median value is close to 1, so we cannot prefer a particular model. Taking into account the above mentioned, we took the SSA model for magnetic field estimation.

---

[3] `https://emcee.readthedocs.io/`
[4] `https://corner.readthedocs.io/`
[5] `https://johannesbuchner.github.io/UltraNest/`

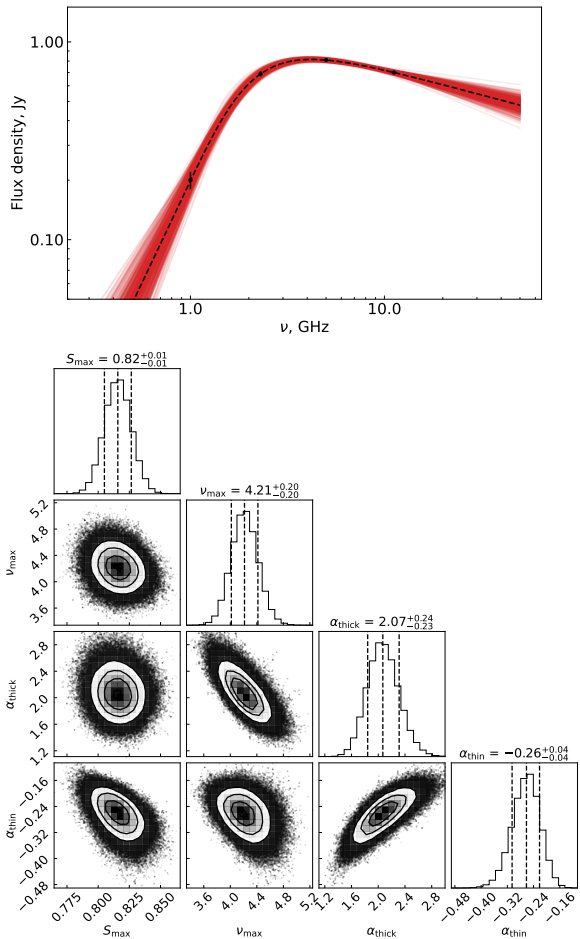

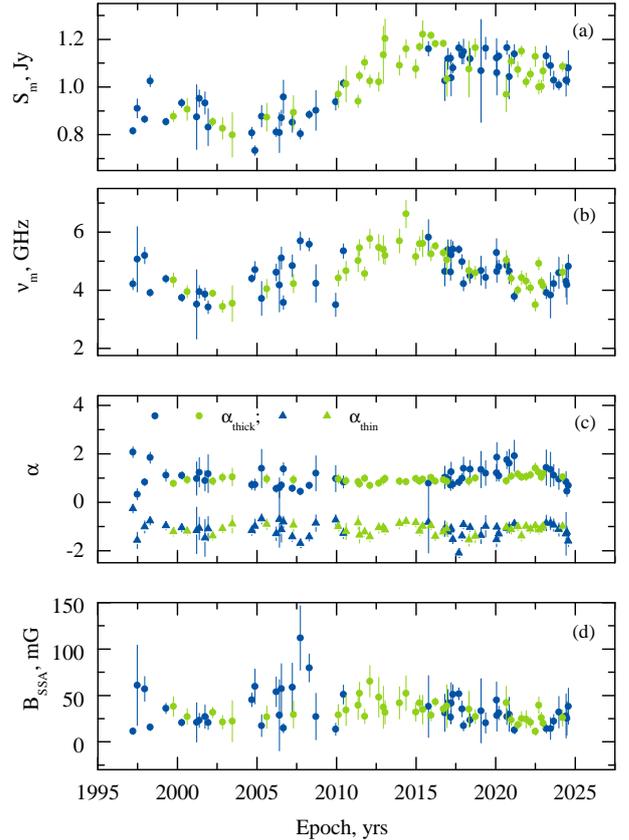

**Figure 9**: An example of MCMC fitting (a spectrum at the epoch 1997.21). Top panel: the red thin lines are the spectra constructed based on the parameter samples from the posterior probability distribution; the dashed black line is the best fit according to the medians of the marginal posteriors. Bottom: the corresponding corner diagrams; the diagonals show the marginal posterior probability distributions, in the other cells 2D projections of the total posterior probability distribution are depicted.

**Figure 10**: Variations of the SSA parameters. The blues symbols present the parameters of $B_{\text{SSA}}$, $\nu_m$, $\alpha_{thick}$, and $\alpha_{thin}$ which were calculated based on the observed quasi-simultaneous spectra from RATAN-600, RT-32, and RT-22; the green symbols show these parameters obtained using the median RATAN-600 data at 1 and 2 GHz.

### 4.4. SSA Magnetic Field Strength

We estimated the magnetic field strength $B_{\text{SSA}}$ (G) assuming that the spectral peak is caused by the SSA processes (Marscher 1983, Slish 1963, Verschuur et al. 1974):

$$B_{\text{SSA}} \approx 10^{-5} b(\alpha_{\text{thin}}) \nu_m^5 \theta^4 S_m^{-2} \delta (1+z)^{-1}, \quad (12)$$

where $b(\alpha_{\text{thin}})$ is a dimensionless parameter, which we calculated according to equation (A14) from Pushkarev et al. (2019), $\nu_m$ is the peak frequency (GHz), $\theta$ is the angular size of the emitting region (mas), $S_m$ is the peak flux density (Jy) derived from the model fitting described above, $\delta$ is the Doppler factor, and $z$

The general results of quasi-simultaneous spectra modeling are shown in Fig. 8 and Fig. 10. The former demonstrates the fitted spectra along with the archival CATS data, and the latter shows variation of the SSA parameters with time.

is the redshift of the source. We took $S_m$, $\nu_m$, and $\alpha_{\text{thin}}$ from the results of SSA fitting (Table 10). The peak frequency $\nu_m$ lies in the range of 4–6 GHz; however, there are only individual data about the angular sizes at 5 GHz (Koryukova et al. 2022, Pushkarev and Kovalev 2015). (Pushkarev and Kovalev 2012) PKS 1614+051 has core sizes $\theta_{2.3} = 1.42$ mas at 2.3 GHz and $\theta_{8.6} = 0.67$ mas at 8.6 GHz (Pushkarev and Kovalev 2012). We can estimate the source's angular size $\theta$ based on the relation $\theta \sim \nu^{-k}$. The coefficient $k \approx -0.575$, according to the data from Pushkarev and Kovalev 2012. As a result, we have angular sizes varying from 1.03 mas to 0.82 mas at $\nu_m = 4$–6 GHz, for example, $\theta_5 = 0.91$ mas at 5 GHz. We estimated the Doppler factor following the approach of Lähteenmäki and Valtaoja (1999) as $\delta = \left(\dfrac{T_{b,\text{var}}}{T_{b,\text{int}}}\right)^{1/3}$, where $T_{b,\text{int}}$ is the intrinsic brightness temperature and $T_{b,\text{var}}$ is the observed variability brightness temperature. We used the equipartition value $5 \times 10^{10}$ K (Readhead 1994) for $T_{b,\text{int}}$, while calculating $T_{b,\text{var}}$ from the radio light curve as in (Lähteenmäki and Valtaoja 1999): $T_{b,\text{var}} = 5.87 \times 10^{21} h^{-2} \dfrac{\lambda^2 S_{\max}}{\tau_{\text{obs}}^2} \left(\sqrt{1+z} - 1\right)^2$, where $\lambda$ is the wavelength in meters, $S_{\max}$ is the outburst amplitude in Jy, $h = H_0/100$ is the normalised Hubble constant, and $\tau_{\text{obs}}$ is the observed variability timescale in days. We get $\delta \approx 2$ under these assumptions. The calculated $B_{\text{SSA}}$ values are presented in Table 10. As mentioned above, we do not have VLBI data to calculate the angular sizes $\theta$ and Doppler factors $\delta$ for the entire set of quasi-simultaneous spectra. Therefore the uncertainties of the magnetic field values are given without considering these factors. The temporary evolution of the magnetic field $B_{\text{SSA}}$ is shown in Fig. 10 (blue points). We see an increasing $B_{\text{SSA}}$ magnetic field in the period of 2005–2010, which then decreases after 2015. Unfortunately, we do not have measurements at 1 and 2 GHz during 2010–2015 to model quasi-simultaneous radio spectra according to the approach described in Section 4.4.3. To overcome this problem, we used the median flux densities at 1 and 2 GHz according to the RATAN-600 measurements over the entire period. This is justified by the fact that at these frequencies the flux density does not show any trend in its light curves (Fig. 1) and the literature data from the CATS database have close flux densities at 1–2 GHz (Fig. 8). Therefore we can consider the radio spectrum at the decimeter range as constant. Under these assumptions we modelled the quasi-simultaneous radio spectra where the measurements had been absent at 1–2 GHz. As a result, we obtained 40 additional estimates of the $B_{\text{SSA}}$ magnetic field (green points in Fig. 10), and 13 of them belong to the 2010–2015 period. We note that the blue and green points correspond well to each other. Also we see a hint of the magnetic field trend to larger values in 2010–2015, reaching its maximum in about 2012–2013, and another surge of $B_{\text{SSA}}$ in 2006–2007. These periods of maximum magnetic field are located before the development of the biggest flare in the light curves (Fig. 1). This fact can be considered as evidence that the flaring power is provided by accumulating magnetic energy (e.g. Sikora and Begelman 2013). The mechanism of magnetic energy accumulation is obviously related to accretion processes. The following output power in the flare can be related to magnetic reconnection.

Table 10: Parameters of the SSA fitting

| Date | $S_m$, | $\nu_m$, | $\alpha_{\text{thick}}$ | $\alpha_{\text{thin}}$ | $B_{\text{SSA}}$, |
|---|---|---|---|---|---|
| yyyy.yy | Jy | GHz | | | mG |
| (1) | (2) | (3) | (4) | (5) | (6) |
| 1997.21 | $0.82^{+0.01}_{-0.01}$ | $4.2^{+0.2}_{-0.2}$ | $2.1^{+0.2}_{-0.2}$ | $-0.3^{+0.1}_{-0.1}$ | $12^{+3}_{-3}$ |
| 1997.47 | $0.91^{+0.04}_{-0.04}$ | $5.1^{+0.7}_{-1.1}$ | $0.3^{+0.4}_{-0.2}$ | $-1.6^{+0.3}_{-0.3}$ | $61^{+43}_{-68}$ |
| 1997.94 | $0.87^{+0.01}_{-0.01}$ | $5.2^{+0.2}_{-0.3}$ | $0.8^{+0.4}_{-0.2}$ | $-1.0^{+0.2}_{-0.2}$ | $57^{+13}_{-16}$ |
| 1998.29 | $1.03^{+0.03}_{-0.02}$ | $3.9^{+0.1}_{-0.1}$ | $1.8^{+0.3}_{-0.2}$ | $-0.8^{+0.1}_{-0.1}$ | $16^{+3}_{-3}$ |
| 1999.28 | $0.85^{+0.01}_{-0.01}$ | $4.4^{+0.1}_{-0.1}$ | $1.1^{+0.4}_{-0.2}$ | $-1.0^{+0.1}_{-0.2}$ | $36^{+6}_{-6}$ |
| 2000.27 | $0.93^{+0.02}_{-0.02}$ | $3.7^{+0.1}_{-0.1}$ | $1.1^{+0.2}_{-0.1}$ | $-1.0^{+0.1}_{-0.1}$ | $21^{+4}_{-4}$ |
| 2001.21 | $0.87^{+0.24}_{-0.14}$ | $3.5^{+0.6}_{-1.2}$ | $1.0^{+1.2}_{-0.7}$ | $-1.2^{+0.8}_{-1.0}$ | $21^{+21}_{-36}$ |
| 2001.38 | $0.95^{+0.04}_{-0.04}$ | $3.9^{+0.2}_{-0.2}$ | $1.2^{+1.0}_{-0.6}$ | $-1.1^{+0.2}_{-0.5}$ | $23^{+8}_{-7}$ |
| 2001.73 | $0.93^{+0.06}_{-0.05}$ | $3.9^{+0.3}_{-0.2}$ | $0.9^{+1.1}_{-0.4}$ | $-1.5^{+0.4}_{-0.8}$ | $27^{+13}_{-8}$ |

Table 10: Parameters of the SSA fitting

| Date | $S_m$, | $\nu_m$, | $\alpha_{\text{thick}}$ | $\alpha_{\text{thin}}$ | $B_{\text{SSA}}$, |
|---|---|---|---|---|---|
| yyyy.yy | Jy | GHz | | | mG |
| (1) | (2) | (3) | (4) | (5) | (6) |
| 2001.93 | $0.83^{+0.09}_{-0.08}$ | $3.4^{+0.2}_{-0.2}$ | $1.2^{+1.1}_{-0.8}$ | $-1.1^{+0.3}_{-0.5}$ | $21^{+7}_{-8}$ |
| 2004.67 | $0.81^{+0.03}_{-0.02}$ | $4.4^{+0.1}_{-0.1}$ | $0.7^{+0.3}_{-0.2}$ | $-1.2^{+0.2}_{-0.2}$ | $45^{+8}_{-7}$ |
| 2004.87 | $0.73^{+0.02}_{-0.02}$ | $4.7^{+0.3}_{-0.3}$ | $0.7^{+0.5}_{-0.2}$ | $-1.0^{+0.3}_{-0.3}$ | $60^{+19}_{-19}$ |
| 2005.29 | $0.88^{+0.05}_{-0.05}$ | $3.7^{+0.5}_{-0.6}$ | $1.4^{+1.0}_{-0.8}$ | $-0.7^{+0.1}_{-0.2}$ | $17^{+12}_{-14}$ |
| 2006.22 | $0.81^{+0.02}_{-0.02}$ | $4.6^{+0.3}_{-0.3}$ | $0.6^{+0.2}_{-0.1}$ | $-1.3^{+0.3}_{-0.3}$ | $54^{+16}_{-17}$ |
| 2006.42 | $0.81^{+0.09}_{-0.08}$ | $4.2^{+1.1}_{-0.9}$ | $0.6^{+1.2}_{-0.4}$ | $-0.7^{+0.4}_{-1.1}$ | $29^{+38}_{-32}$ |
| 2006.53 | $0.87^{+0.03}_{-0.03}$ | $5.1^{+0.5}_{-0.4}$ | $0.7^{+0.3}_{-0.2}$ | $-1.1^{+0.3}_{-0.5}$ | $57^{+27}_{-22}$ |
| 2006.66 | $0.96^{+0.09}_{-0.07}$ | $3.6^{+0.2}_{-0.2}$ | $1.4^{+0.3}_{-0.3}$ | $-0.8^{+0.2}_{-0.2}$ | $15^{+5}_{-5}$ |
| 2007.22 | $0.85^{+0.04}_{-0.04}$ | $4.8^{+0.4}_{-0.4}$ | $0.6^{+0.1}_{-0.1}$ | $-1.4^{+0.2}_{-0.2}$ | $59^{+26}_{-24}$ |
| 2007.73 | $0.80^{+0.02}_{-0.02}$ | $5.7^{+0.3}_{-0.3}$ | $0.4^{+0.1}_{-0.1}$ | $-1.7^{+0.1}_{-0.2}$ | $112^{+35}_{-31}$ |
| 2008.30 | $0.88^{+0.02}_{-0.02}$ | $5.6^{+0.2}_{-0.2}$ | $0.7^{+0.1}_{-0.2}$ | $-1.4^{+0.1}_{-0.1}$ | $80^{+15}_{-16}$ |
| 2008.70 | $0.90^{+0.10}_{-0.08}$ | $4.2^{+0.8}_{-0.6}$ | $1.2^{+1.0}_{-0.7}$ | $-0.9^{+0.2}_{-0.3}$ | $27^{+25}_{-21}$ |
| 2009.95 | $0.94^{+0.04}_{-0.03}$ | $3.5^{+0.3}_{-0.4}$ | $1.0^{+0.9}_{-0.5}$ | $-0.7^{+0.1}_{-0.2}$ | $14^{+7}_{-8}$ |
| 2010.44 | $1.02^{+0.02}_{-0.02}$ | $5.3^{+0.2}_{-0.2}$ | $0.8^{+0.3}_{-0.2}$ | $-1.3^{+0.2}_{-0.2}$ | $51^{+10}_{-12}$ |
| 2015.78 | $1.16^{+0.05}_{-0.03}$ | $5.8^{+1.0}_{-0.6}$ | $0.8^{+1.0}_{-0.3}$ | $-0.8^{+0.6}_{-1.3}$ | $38^{+33}_{-20}$ |
| 2016.81 | $1.03^{+0.09}_{-0.08}$ | $4.6^{+0.6}_{-0.5}$ | $1.0^{+0.8}_{-0.5}$ | $-1.1^{+0.2}_{-0.3}$ | $31^{+20}_{-18}$ |
| 2017.01 | $1.12^{+0.05}_{-0.04}$ | $5.4^{+0.4}_{-0.3}$ | $1.0^{+0.4}_{-0.3}$ | $-1.2^{+0.1}_{-0.2}$ | $42^{+15}_{-14}$ |
| 2017.18 | $1.12^{+0.02}_{-0.02}$ | $4.6^{+0.2}_{-0.3}$ | $0.7^{+0.3}_{-0.2}$ | $-1.1^{+0.1}_{-0.1}$ | $27^{+7}_{-8}$ |
| 2017.22 | $1.04^{+0.08}_{-0.08}$ | $5.2^{+0.5}_{-0.5}$ | $1.2^{+0.6}_{-0.4}$ | $-1.1^{+0.2}_{-0.2}$ | $42^{+22}_{-22}$ |
| 2017.32 | $1.08^{+0.02}_{-0.02}$ | $5.4^{+0.2}_{-0.3}$ | $0.7^{+0.2}_{-0.2}$ | $-1.5^{+0.1}_{-0.1}$ | $51^{+10}_{-13}$ |
| 2017.70 | $1.16^{+0.02}_{-0.02}$ | $5.4^{+0.1}_{-0.1}$ | $0.8^{+0.2}_{-0.1}$ | $-2.1^{+0.2}_{-0.2}$ | $52^{+6}_{-7}$ |
| 2017.89 | $1.13^{+0.04}_{-0.04}$ | $5.0^{+0.2}_{-0.2}$ | $1.0^{+0.3}_{-0.2}$ | $-1.4^{+0.2}_{-0.2}$ | $35^{+9}_{-9}$ |
| 2017.98 | $1.15^{+0.06}_{-0.05}$ | $4.2^{+0.2}_{-0.2}$ | $1.4^{+0.7}_{-0.5}$ | $-0.9^{+0.1}_{-0.2}$ | $17^{+6}_{-5}$ |
| 2018.39 | $1.12^{+0.05}_{-0.04}$ | $4.5^{+0.3}_{-0.2}$ | $1.4^{+0.5}_{-0.4}$ | $-1.0^{+0.1}_{-0.2}$ | $23^{+7}_{-7}$ |
| 2019.08 | $1.07^{+0.27}_{-0.22}$ | $4.7^{+0.8}_{-0.5}$ | $1.3^{+1.0}_{-0.8}$ | $-1.4^{+0.2}_{-0.3}$ | $33^{+35}_{-22}$ |
| 2019.37 | $1.16^{+0.05}_{-0.05}$ | $4.4^{+0.5}_{-0.4}$ | $1.2^{+0.9}_{-0.5}$ | $-1.0^{+0.2}_{-0.4}$ | $20^{+11}_{-9}$ |
| 2020.07 | $1.12^{+0.08}_{-0.08}$ | $5.3^{+0.4}_{-0.5}$ | $1.2^{+0.5}_{-0.3}$ | $-1.5^{+0.3}_{-0.3}$ | $45^{+20}_{-22}$ |

| Date | $S_m$, | $\nu_m$, | $\alpha_{\text{thick}}$ | $\alpha_{\text{thin}}$ | $B_{\text{SSA}}$, |
|---|---|---|---|---|---|
| yyyy.yy | Jy | GHz | | | mG |
| (1) | (2) | (3) | (4) | (5) | (6) |
| 2020.08 | $1.06^{+0.10}_{-0.09}$ | $4.6^{+0.5}_{-0.4}$ | $1.9^{+0.7}_{-0.6}$ | $-1.0^{+0.1}_{-0.2}$ | $28^{+17}_{-13}$ |
| 2020.21 | $1.13^{+0.01}_{-0.01}$ | $4.8^{+0.2}_{-0.2}$ | $1.1^{+0.3}_{-0.2}$ | $-1.3^{+0.1}_{-0.1}$ | $31^{+8}_{-8}$ |
| 2020.72 | $1.17^{+0.03}_{-0.03}$ | $4.8^{+0.3}_{-0.3}$ | $1.7^{+0.5}_{-0.4}$ | $-1.1^{+0.2}_{-0.2}$ | $27^{+9}_{-9}$ |
| 2020.86 | $1.04^{+0.10}_{-0.09}$ | $4.6^{+0.5}_{-0.4}$ | $1.6^{+0.8}_{-0.6}$ | $-1.1^{+0.2}_{-0.2}$ | $30^{+19}_{-15}$ |
| 2021.19 | $1.14^{+0.04}_{-0.04}$ | $3.8^{+0.2}_{-0.2}$ | $1.9^{+0.7}_{-0.6}$ | $-0.9^{+0.1}_{-0.1}$ | $13^{+4}_{-3}$ |
| 2023.19 | $1.13^{+0.05}_{-0.04}$ | $3.9^{+0.2}_{-0.2}$ | $1.4^{+1.0}_{-0.7}$ | $-0.9^{+0.1}_{-0.3}$ | $14^{+5}_{-4}$ |
| 2023.45 | $1.09^{+0.07}_{-0.04}$ | $3.8^{+0.6}_{-0.8}$ | $1.4^{+0.9}_{-0.7}$ | $-0.9^{+0.2}_{-0.3}$ | $14^{+12}_{-15}$ |
| 2023.66 | $1.03^{+0.04}_{-0.04}$ | $4.2^{+0.3}_{-0.3}$ | $1.1^{+0.8}_{-0.4}$ | $-0.9^{+0.2}_{-0.3}$ | $22^{+7}_{-7}$ |
| 2023.98 | $1.01^{+0.02}_{-0.02}$ | $4.6^{+0.5}_{-0.5}$ | $1.0^{+0.6}_{-0.4}$ | $-1.1^{+0.2}_{-0.3}$ | $32^{+17}_{-19}$ |
| 2024.43 | $1.03^{+0.08}_{-0.07}$ | $4.3^{+0.7}_{-0.5}$ | $0.8^{+1.1}_{-0.4}$ | $-1.3^{+0.5}_{-0.9}$ | $28^{+24}_{-18}$ |
| 2024.49 | $1.03^{+0.02}_{-0.02}$ | $4.2^{+0.6}_{-0.7}$ | $0.5^{+0.3}_{-0.2}$ | $-1.2^{+0.2}_{-0.2}$ | $25^{+18}_{-20}$ |
| 2024.57 | $1.08^{+0.07}_{-0.07}$ | $4.8^{+0.5}_{-0.4}$ | $0.7^{+0.1}_{-0.1}$ | $-1.6^{+0.2}_{-0.2}$ | $38^{+19}_{-17}$ |

## 5. OPTICAL SPECTROSCOPY

The slit of the SCORPIO-I spectrograph was rotated to a position angle $\sim 35°$ to pass through the centers of PKS 1614+051 and a companion galaxy revealed and studied by Husband et al. (2015) in the course of the mVLT Multi-Object Spectroscopic Explorer (MUSE) science verification observations. The total spectra of the blazar and the companion galaxy are presented in Fig. 11.

A detailed inspection of individual spectra of the central and neighboring regions of the blazar shows possible variation of the Ly$\alpha$ profile, as it can be seen in Fig. 12 (upper panel). In order to investigate these spatial variation, we approximated the individual spectra integrated with a $1''\!.8$ step along the slit by a composition of two Gaussians responsible for the emission and absorption components, respectively. To confirm our results of the emission line Gaussian decomposition, we also use the spectral data for PKS 1614+051 taken with SCORPIO at another epoch: on the night of 30.06/01.07.2024 with a better spectral resolution but with a somewhat worse signal-to-noise ratio. The slit was oriented at a position angle $\sim 40°$. The result is presented in Fig. 12 (lower panel).

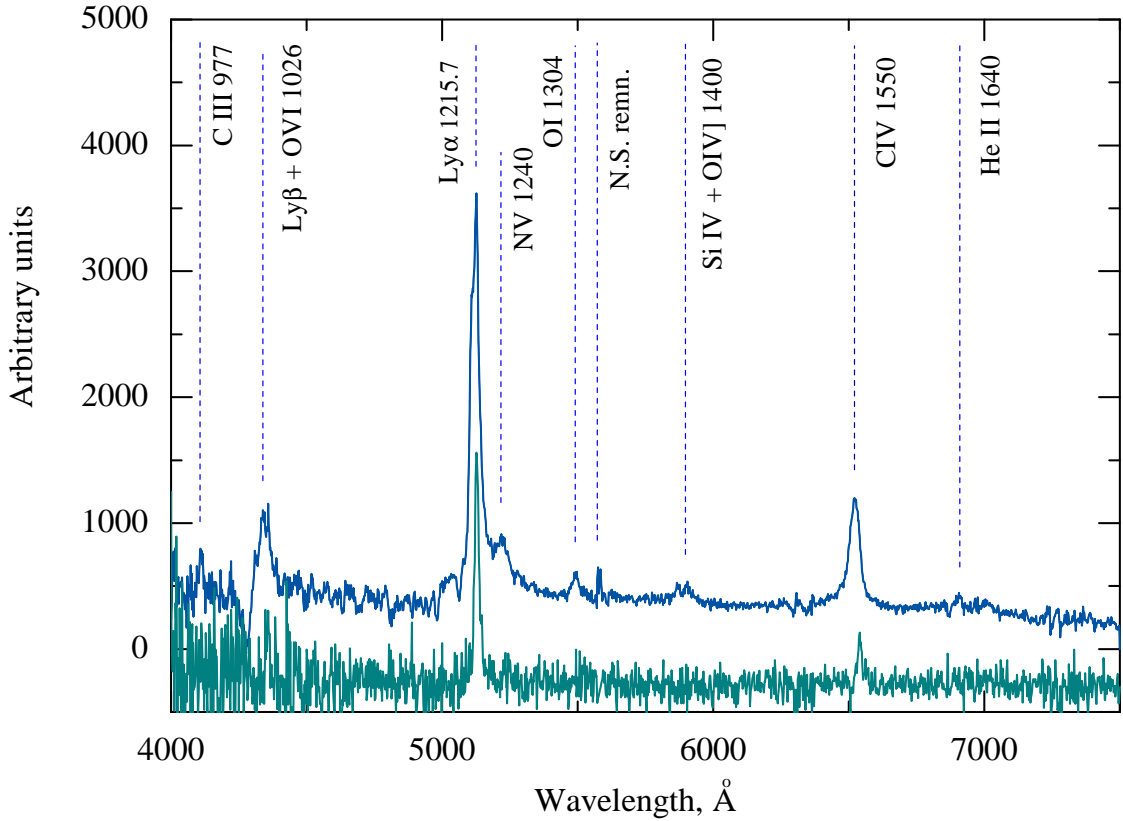

**Figure 11**: The optical spectra of PKS 1614+051 (blue color) and a companion Lyα-emitting galaxy (green line). The data are presented in arbitrary energetic units. For convenience, the data for the Lyα galaxy were multiplied by a factor of 5. The most prominent spectral lines are marked by vertical dashed lines.

The results of profile approximation are shown in Fig. 13, where the redshift of Lyα components is shown as a function of the coordinate along the slit. The red squares show the redshift of the Lyα emission for three regions around the center (distance = 0) and two measurements for the companion AGN. The green and blue squares mark the redshift of the absorption components for the 14/15.07.2024 and 30.06/01.07.2024 nights, respectively, which are clearly seen in the profiles of the Lyα emission line. The magenta squares show the accuracy of the positional measurements for the night sky $\lambda\,5577$ Å line. The results for the emission component indicate a significant difference ($\sim 300$ km s$^{-1}$ in the rest frame) in the radial velocities between the blazar nucleus and the companion galaxy, which is in good agreement with the results of previous studies (Djorgovski et al. 1985, Husband et al. 2015). Because of the exposure time of only 30 min, we could not detect any signs of the Lyα bridge, which was studied in detail with MUSE in the latter paper.

## 6. SUMMARY AND CONCLUSIONS

The nature of HFPs has not been yet completely understood, and the high-redshift blazar PKS 1614+051 is a useful probe to study the early stage of AGN evolution. Our analysis of the PKS 1614+051 long-term variability reveals that it is quite marginal (10%–20%) at the peak frequency and around it, and the light curve shows rather smooth variation at these frequencies. The position of the spectral peak is relatively stable, it has a normal distribution with a median value of 4.6 GHz during its 27-yr monitoring. The 5–37 GHz light curves correlate having progressive observed delays from 0.3 to 6.4 years with decreasing radio frequency, which can be explained both in terms of the SSA mechanism with different opacity at the low and high frequencies and by different sites of particle acceleration at the observed frequencies (e.g., Marscher and Gear 1985).

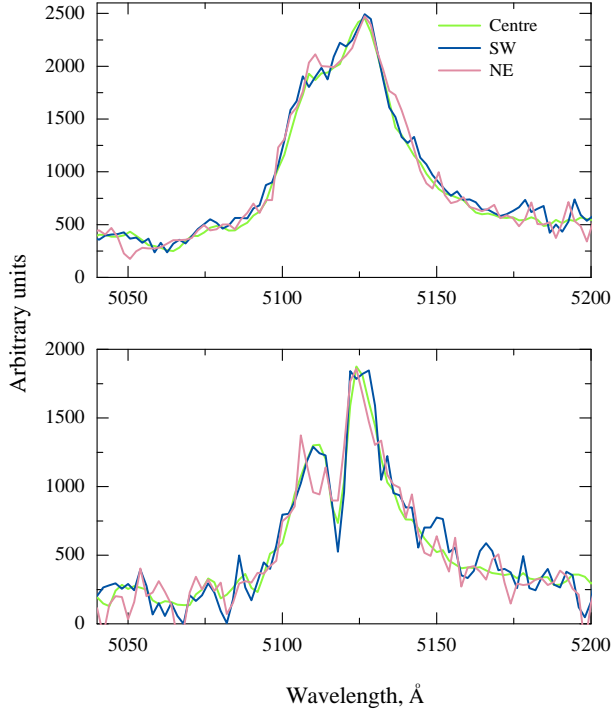

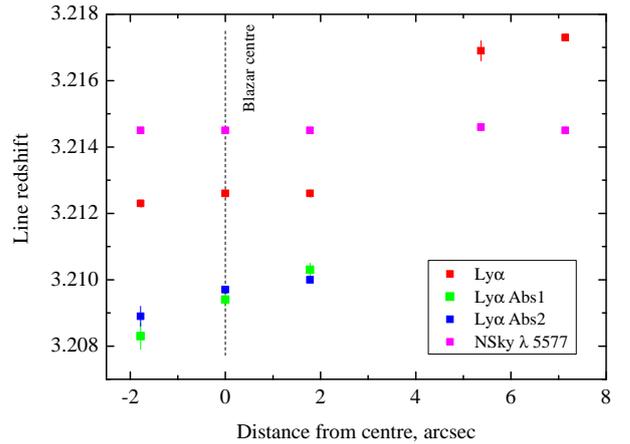

**Figure 12**: Top panel: three subsequent spectra of the central region of PKS 1614+051, integrated with a 1″.8 step along the spectrograph slit, taken on the night of 14/15.07.2024. Bottom panel: the spectra taken with SCORPIO-I on the night of 30.06/01.07.2024. The spectra are presented with different colors and normalized to the same maximum for better comparison.

**Figure 13**: The redshift of the Lyα line components: the wide emission and narrow absorption lines. The night sky [OI] λ 5577 Å line is shown for control. The coordinate 0 corresponds to the blazar center position, the abscissa increases in the direction to the companion galaxy, which is located at a distance of 5″.4–7″.1 from the center.

The relation "observed lag vs frequency" is a useful tool to estimate how fast the flare disturbance propagates down the jet (see, e.g., Krishna Mohana et al. 2024). Since PKS 1614+051 is extremely compact and there are no measurements of its jet kinematics, such estimates would be useful to understand the parsec jet behavior in this blazar. However, the presented DCF analysis has not found a clear picture of how the lag changes with the frequency. Using three frequencies—5, 8 and 11 GHz—we can make just a rough calculation: the lag versus frequency plot is fitted by straight lines with a negative slope, on average as $\sim -0.6$ yr GHz$^{-1}$. The found negative slope supports the scenarios where time lags at lower frequencies are due to greater opacities, as greater SSA is observed at the lower frequencies compared to the higher ones.

The PKS 1614+051 variability timescales of 0.2–1.8 yrs (rest frame) at the radio frequencies are comparable with the variability scale of the blazars at the intermediate and low redshifts. This means that for such distant blazars a greater monitoring timescale is needed in the observer's frame of reference to determine their variability properties. The established low variability level for many distant HFPs can be caused by the limited monitoring time in the rest frame.

The daily observations at 5 GHz in 2019–2020

reveal a low radio variability level $F_{\mathrm{var}} = 0.02$ and $M = 0.03$ and the shortest variability time scale $\tau_{\mathrm{rest}} = 12$ days. However, we have found that the influence of RISS is significant and comparable to the flux density modulation index.

The VLBA maps at 15 GHz show that PKS 1614+051 has two components, which are separated by $\sim 1.4$ mas and have a flux density ratio $\sim 3.5$ (Orienti et al. 2006a), suggesting the core–jet morphology on the parsec scale. We estimated the size of the emission region at 8 GHz as:

$$R \leq c \cdot t_{\mathrm{var}} \cdot \delta/(1+z), \qquad (13)$$

assuming the redshift $z = 3.21$, the variability timescale $t_{\mathrm{var}} = 6.9$ yrs, and $\delta = 3$ as the Doppler factor. For the latter parameter we adopted the equipartition brightness temperature $T_{\mathrm{eq}} \simeq 5 \times 10^{10}$ K suggested by Readhead (1994) and estimated $\delta = T_b/T_{\mathrm{eq}}$, where $T_b = 1.83 \times 10^{11}$ K (Pushkarev and Kovalev 2012). The derived $R < 1.3$ pc is about two times more compact than the VLBI core size at 8.6 GHz ($\sim 2$ pc) that we estimated from its angular size in Pushkarev and Kovalev (2012) and Koryukova et al. (2022) ($\theta \sim 0.264 \pm 0.002$ mas, Fig. 14).

Thus, the combination of the PKS 1614+051 variability features and the permanent convex spectral shape could be explained by the emission from the dominant Northern component, diluted by the variable fainter Southern component. The high brightness temperatures of the source measured at 2.3 and 8.6 GHz as $\sim 8 \times 10^{11}$ K and $\sim 2 \times 10^{11}$ K in Pushkarev and Kovalev (2012) also support the Doppler-boosted synchrotron emission scenario typical of the beamed objects as well as the high fraction ($> 2\%$) of polarized emission (Orienti 2007). Thus, our results based on the new observed data provide support for the blazar nature of the HFP source PKS 1614+051.

As we have found from the results of the spectroscopic study with the BTA, the position of the narrow absorption lines in the Ly$\alpha$ profiles (their widths in our data do not differ from the instrumental contour width: about 650 and 350 km s$^{-1}$ for the two nights) shows a systematic trend along the slit from the SW to the NE around the blazar center: about $\pm 90$ and $\pm 45$ km s$^{-1}$ in the rest frame in both directions for different slit orientations. These values are similar within their measurement accuracy. As Husband et al. (2015) concluded, the presence of this feature may be an indicator of a large neutral gas cloud or a halo with a radius of $\geq 50$ kpc. Our data may indicate the existence of a gaseous disk with a definite rotation direction instead of random motions with some velocity dispersion. The presence of gas in the torus or disk structures was proposed by Orienti et al. (2006b) to explain the observed phenomena in extremely young radio galaxies.

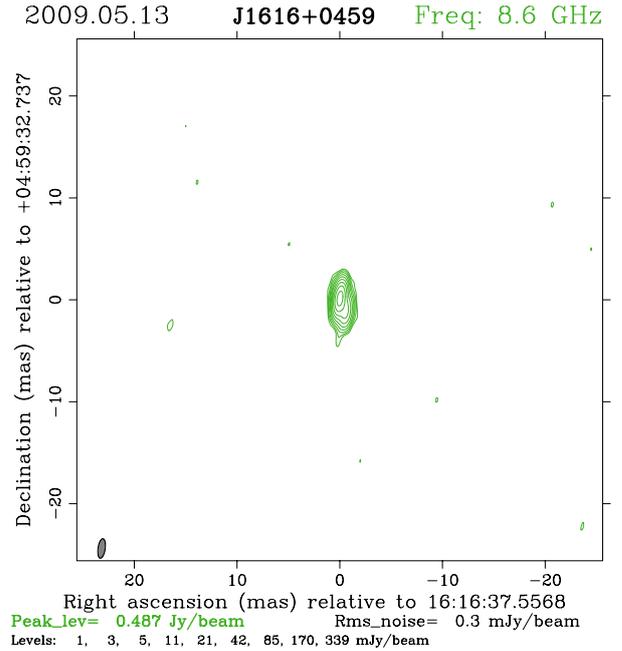

**Figure 14**: The X-band VLBI map of PKS 1614+051 derived from the atsrogeo database for the epoch May 13, 2009.

Additional observations with another slit orientation should be made to confirm any hypotheses about neutral gas dynamics and, possibly, make clear the model of its origin. Thus, this result does not contradict the opinion about PKS 1614+051 as of a young object at the stage of star formation from an extended dense gaseous envelope: the gas regularly rotates around the active nucleus and produces a very strong absorption feature, which is clearly seen in high-resolution spectra (Bechtold 1994).

The optical study with the BTA SCORPIO-

I spectrograph allows us to suspect the presence of the gaseous component regular motion around the center of the blazar. This phenomenon may support the idea about the young age of PKS 1614+051, as it follows from its radio characteristics.

As it has been mentioned above, we cannot prefer the SSA model to the inhomogeneous FFA model from the statistical point of view. One of the reasons may be a scarcity of measurement points in the decimeter range for a more detailed modelling of the quasi-simultaneous spectra. The other possible reason is the comparable contribution of both processes, the SSA and FFA, into the resulting radio spectrum. We suggest that the results of our BTA+SCORPIO-I spectroscopic observations confirm the presence of enough gaseous matter needed to form an external FFA screen. The equipartition magnetic field $B_{\rm eq}$ (e.g., Miley 1980) for PKS 1614+051 is 150–170 mG and significantly exceeds $B_{\rm SSA}$. Recalling all the mentioned assumptions about the estimation of the magnetic field, we cannot confidently conclude about the difference for these quantities; nevertheless, this result can be considered as a hint about the co-existing SSA and FFA processes in PKS 1614+051. Therefore, we conclude that it is necessary to develop complex absorption models allowing for both the SSA as well as FFA processes to describe the peaked radio spectra of PKS 1614+051 and other distant quasars. This is an aim for future work.

ACKNOWLEDGEMENTS

The observations were carried out with the RATAN-600 scientific facility, the BTA, Zeiss-1000, and AS-500/2 optical reflectors of SAO RAS, and the RT-22 of CrAO RAS. The observations at 5.05 and 8.63 GHz were performed with the Badary and Zelenchukskaya RT-32 radio telescopes operated by the Shared Research Facility Center for the Quasar VLBI Network of IAA RAS (https://iaaras.ru/cu-center/).

YYK was supported by the M2FINDERS project which has received funding from the European Research Council (ERC) under the European Union's Horizon2020 Research and Innovation Programme (grant agreement No. 101018682).

VAE and VLN are grateful to the staff of the Radio Astronomy Department of CrAO RAS for their participation in the observations. VVV thanks Alexander Vinokurov and Roman Uklein from SAO RAS staff for assistance with the SCORPIO-I observations.

This research has made use of the NASA/IPAC Extragalactic Database (NED), which is operated by the Jet Propulsion Laboratory, California Institute of Technology, under contract with the National Aeronautics and Space Administration; the CATS database, available on the Special Astrophysical Observatory website; the SIMBAD database, operated at CDS, Strasbourg, France. This research has made use of the VizieR catalogue access tool, CDS, Strasbourg, France. We used in our work the Astrogeo VLBI FITS image database, DOI: 10.25966/kyy8-yp57, maintained by Leonid Petrov.

FUNDING

The reported study was funded by the Ministry of Science and Higher Education of the Russian Federation under contract 075-15-2022-1227.

CONFLICT OF INTEREST

The authors of this work declare that they have no conflicts of interest.

DATA AVAILABILITY

The data underlying this article are available in the article and in its online supplementary material. The new measured flux densities are distributed in the VizieR Information System. The archival radio data are available in the CATS database on https://www.sao.ru/cats/. The RATAN-600 data are partly available in the BLcat on-line catalogue on https://www.sao.ru/blcat/.